\documentclass[a4paper]{jpconf}
\usepackage{graphicx}
\begin{document}

\def\to{\rightarrow}
\def\gev{\mbox{GeV}}
\def\ev{\mbox{eV}}
\def\mev{\mbox{MeV}}
\def\tev{\mbox{TeV}}
\def\cm{\mbox{cm}}
\def\mpc{\mbox{Mpc}}
\def\AJ{{\it Astroph. J.} }
\def\AJL{{\it Ap. J. Lett.} }
\def\AJS{{\it Ap. J. Supp.} } 
\def\AM{{\it Ann. Math.} }
\def\AP{{\it Ann. Phys.} }
\def\APJ{{\it Ap. J.} }
\def\APP{{\it Acta Phys. Pol.} }
\def\ASJ{{\it Astron. J.} }
\def\ASAS{{\it Astron. and Astrophys.} }
\def\BAMS{{\it Bull. Am. Math. Soc.} }
\def\CMJ{{\it Czech. Math. J.} }
\def\CMP{{\it Commun. Math. Phys.} }
\def\CTP{{\it Commun. Theor. Phys.} }
\def\CQG{{\it Class. Quantum Gravity} }
\def\EURO{{\it Eur. Phys. J.} }
\def\FP{{\it Fortschr. Physik} }
\def\GRG{{\it Gen. Relativity and Gravitation} }
\def\HPA{{\it Helv. Phys. Acta} }
\def\IJMP{{\it Int. J. Mod. Phys.} }
\def\JMM{{\it J. Math. Mech.} }
\def\JHEP{{\it JHEP} }
\def\JP{{\it J. Phys.} }
\def\JCP{{\it J. Chem. Phys.} }
\def\LNC{{\it Lett. Nuovo Cimento} }
\def\SNC{{\it Suppl. Nuovo Cimento} }
\def\MPL{{\it Mod. Phys. Lett.} }
\def\MNRAS{{\it Mon. Not. R. Ast. Soc.} }
\def\NAT{{\it Nature} }
\def\NAST{{\it New Astronomy}} 
\def\NC{{\it Il Nuovo Cimento} }
\def\NP{{\it Nucl. Phys.} }
\def\PL{{\it Phys. Lett.} }
\def\PR{{\it Phys. Rev.} }
\def\PRL{{\it Phys. Rev. Lett.} }
\def\PRTS{{\it Physics Reports} }
\def\PS{{\it Physica Scripta} }
\def\PTP{{\it Progr. Theor. Phys.} }
\def\RMPA{{\it Rev. Math. Pure Appl.} }
\def\RMP{{\it Rev. Mod. Phys.} }
\def\RNC{{\it Rivista del Nuovo Cimento} }
\def\SC{{\it Science} }
\def\SJPN{{\it Soviet J. Part. Nucl.} }
\def\SP{{\it Soviet. Phys.} }
\def\TMF{{\it Teor. Mat. Fiz.} }
\def\TMP{{\it Theor. Math. Phys.} }
\def\YF{{\it Yadernaya Fizika} }
\def\ZETF{{\it Zh. Eksp. Teor. Fiz.} }
\def\ZP{{\it Z. Phys.} }
\def\ZMP{{\it Z. Math. Phys.} }

\def\al{\alpha}
\def\be{\beta}
\def\ga{\gamma}
\def\de{\delta}
\def\ep{\epsilon}
\def\ve{\varepsilon}
\def\ze{\zeta}
\def\et{\eta}
\def\th{\theta}
\def\vt{\vartheta}
\def\io{\iota}
\def\ka{\kappa}
\def\la{\lambda}
\def\vpi{\varpi}
\def\rh{\rho}
\def\vr{\varrho}
\def\si{\sigma}
\def\vs{\varsigma}
\def\ta{\tau}
\def\up{\upsilon}
\def\ph{\phi}
\def\vp{\varphi}
\def\ch{\chi}
\def\ps{\psi}
\def\om{\omega}
\def\Ga{\Gamma}
\def\De{\Delta}
\def\Th{\Theta}
\def\La{\Lambda}
\def\Si{\Sigma}
\def\Up{\Upsilon}
\def\Ph{\Phi}
\def\Ps{\Psi}
\def\Om{\Omega}
\def\mn{{\mu\nu}}
\def\cl{{\cal L}}
\def\fr#1#2{{{#1} \over {#2}}}
\def\prt{\partial}
\def\ap{\al^\prime}
\def\apt{\al^{\prime 2}}
\def\apth{\al^{\prime 3}}
\def\pt#1{\phantom{#1}}
\def\vev#1{\langle {#1}\rangle}
\def\bra#1{\langle{#1}|}
\def\ket#1{|{#1}\rangle}
\def\bracket#1#2{\langle{#1}|{#2}\rangle}
\def\expect#1{\langle{#1}\rangle}
\def\sbra#1#2{\,{}_{{}_{#1}}\langle{#2}|}
\def\sket#1#2{|{#1}\rangle_{{}_{#2}}\,}
\def\sbracket#1#2#3#4{\,{}_{{}_{#1}}
 \langle{#2}|{#3}\rangle_{{}_{#4}}\,}
\def\sexpect#1#2#3{\,{}_{{}_{#1}}\langle{#2}\rangle_{{}_{#3}}\,}
\def\half{{\textstyle{1\over 2}}}
\def\frac#1#2{{\textstyle{{#1}\over {#2}}}}
\def\ni{\noindent}
\def\lsim{\mathrel{\rlap{\lower4pt\hbox{\hskip1pt$\sim$}}
    \raise1pt\hbox{$<$}}}
\def\gsim{\mathrel{\rlap{\lower4pt\hbox{\hskip1pt$\sim$}}
    \raise1pt\hbox{$>$}}}
\def\sqr#1#2{{\vcenter{\vbox{\hrule height.#2pt
         \hbox{\vrule width.#2pt height#1pt \kern#1pt
         \vrule width.#2pt}
         \hrule height.#2pt}}}}
\def\square{\mathchoice\sqr66\sqr66\sqr{2.1}3\sqr{1.5}3}
\newcommand{\rf}[1]{(\ref{#1})} 
\def\beq{\begin{equation}}
\def\eeq{\end{equation}}
\def\beqa{\begin{eqnarray}} 
\def\eeqa{\end{eqnarray}}

\def\laq{\raise 0.4 ex \hbox{$<$}\kern -0.8 em\lower 0.62 ex\hbox{$\sim$}}
\def\gaq{\raise 0.4 ex \hbox{$>$}\kern -0.7 em\lower 0.62 ex\hbox{$\sim$}}

\title{The Gravitational Quantum Well}

\author{O. Bertolami\footnote[1]{Speaker}, J.G. Rosa\footnote[2]{Present address: Department of Applied Mathematics and 
Theoretical Physics, University of Cambridge, UK}}

\address{Instituto Superior T\'ecnico, Departamento de
F\'{\i}sica, Av. Rovisco
Pais, 1049-001 Lisboa, Portugal}

\ead{orfeu@cosmos.ist.utl.pt, jpt35@cam.ac.uk}

\begin{abstract}
We discuss the implications of a model of noncommutative Quantum Mechanics 
where noncommutativity is extended to the phase space. 
We analyze how this model affects the problem of the 
two-dimensional gravitational quantum well and use the latest experimental 
results for the energy states of neutrons in the Earth's 
gravitational field to establish an upper bound on the fundamental momentum 
scale introduced by noncommutativity. We show that the 
configuration space noncommutativity has, in leading order, no effect 
on the problem and that in the context of the model, 
a correction to the presently accepted value of Planck's 
constant to $1$ part in $10^{24}$ arises. 

We also study the transition between quantum and classical behaviour 
of particles in a gravitational quantum well and analyze how an increase 
in the particles mass turns the energy spectrum into a continuous one. 
We consider these effects and argue that they could be tested by through 
experiments with atoms and fullerene-type molecules.

\end{abstract}

\section{Introduction}

In this contribution we discuss a model for noncommutative Quantum Mechanics (NCQM) where 
noncommutativity in the phase space is 
considered in the context of the gravitational quantum
well (GQW) \cite{Bertolami_1,Bertolami_2}. Furthermore, we consider the dependence of 
the energy spectrum of the GQW on the mass of the particles in order to study the quantum 
to classical transition \cite{Bertolami_3}

Recently, there has been quite some interest on the noncommutative 
geometry in the context of Quantum Mechanics.
This has its roots on the role that noncommutative geometry plays in unification models and 
in string theory. Indeed, since the discovery that the low-energy effective theory 
of a D-brane in the background of a NS-NS B field lives in a noncommutative 
space \cite{Connes, Seiberg}, many efforts have been devoted to the 
study of noncommutative field theories (see \cite{Bertolami_4} and refs. therein), to the 
understanding of the coupling to gravity 
(see \cite{Bertolami_5} and refs. therein) and, to the non-relativistic limit, through
versions of NCQM \cite{Ho, Nair, Zhang_1, Zhang_2, Demetrian, Gamboa, Li, Bertolami_1, Bertolami_2}.

Although in string theory only the coordinates space exhibits a 
noncommutative structure, models in which the noncommutative geometry is extended to the phase space have been 
recently much discussed  \cite{Zhang_1, Djemai, Bertolami_1}. 
The main argument for this approach is that noncommutativity between momenta arises naturally 
as a consequence of noncommutativity between coordinates, as momenta are defined as the partial derivatives of the action with 
respect to the noncommutative coordinates \cite{Singh}. In a 4-dimensional space, this type of phase space structure is 
defined through the following algebra:
\beqa \label{noncommutation_1}
\lbrack x^{\mu},x^{\nu}]=&i\theta^{\mu\nu}~,\qquad
\lbrack p^{\mu},p^{\nu}]=i\eta^{\mu\nu}~,\qquad
\lbrack x^{\mu},p^{\nu}]=i\hbar\delta^{\mu\nu}~,
\eeqa
where the parameters $\theta^{\mu\nu}$ and $\eta^{\mu\nu}$ are antisymmetric. This algebra is consistent 
with standard Quantum Mechanics thanks the last commutation relation in the set of Eqs. (\ref{noncommutation_1}).

In a recent paper \cite{Bertolami_1}, a 2-dimensional version of this algebra, defined by the noncommutative parameters 
$\theta$ and $\eta$ and by Planck's constant $\hbar$ has been considered
\beqa \label{noncommutation_2}
\lbrack x,y]&=&i\theta~,\qquad
\lbrack p_x,p_y]=i\eta~,\qquad
\lbrack x_i,p_j]=i\hbar\delta_{ij}\qquad i=1,2~.
\eeqa
The implementation of this algebra can be performed by considering the noncommutative variables  $\{x,y,p_x,p_y\}$ 
from the commutative variables $\{x',y',p_x',p_y'\}$ through linear transformations. Two sets of linear transformations can be used
to obtain the complete algebra, Eqs. (\ref{noncommutation_2}). 
The first kind of transformation affects only the variables $x$ and $p_y$, while the second affects 
variables $y$ and $p_x$, thus leading to different results. To eliminate this ambiguity it is convenient to 
consider a set of linear transformations that modify simultaneously all variables:
\beqa \label{linear_1}
x=\zeta\bigg(x'-{\theta\over2\hbar}p_y'\bigg)~,\ \ 
y=\zeta\bigg(y'+{\theta\over2\hbar}p_x'\bigg)~,\ \ 
p_x=\zeta\bigg(p_x'+{\eta\over2\hbar}y'\bigg)~,\ \ 
p_y=\zeta\bigg(p_y'-{\eta\over2\hbar}x'\bigg)~,
\eeqa 
where $\zeta$ is a scaling factor.
In this way, the commutation relations between coordinates and the ones between momenta in Eq. (\ref{noncommutation_2}) are 
recovered, however the commutation relation between coordinates and momenta is altered to
\beq \label{noncommutation_3}
[x_i,p_j]=i\hbar_{eff}\delta_{ij}\qquad i=1,2~, 
\eeq
where we have defined the \emph{effective Planck constant} $\hbar_{eff}=\hbar\big(1+\theta\eta/4\hbar^2\big)$ 
and set $\zeta=1$. 
In a 4-dimensional space, the generalization of the linear transformations Eq. (\ref{linear_1}) may be written as:
\beq \label{linear_2}
x^{\mu}=\zeta\Bigg(x'^{\mu}-{\theta^{\mu}_{\ \nu}\over2\hbar}p'^{\nu}\Bigg)~,\qquad
p^{\mu}=\zeta\Bigg(p'^{\mu}+{\eta^{\mu}_{\ \nu}\over2\hbar}x'^{\nu}\Bigg)~.
\eeq
If one chooses $\zeta=1$, these transformations lead to the following 4-dimensional algebra:
\beqa \label{noncommutation_4}
\lbrack x^{\mu},x^{\nu}]&=&i\theta^{\mu\nu}~,\qquad
\lbrack p^{\mu},p^{\nu}]=i\eta^{\mu\nu}~,\qquad
\lbrack x^{\mu},p^{\nu}]=i\hbar\bigg(\delta^{\mu\nu}+{\theta^{\mu\alpha}\eta^{\nu}_{\ \alpha}\over4\hbar^2}\bigg)~.
\eeqa
Hence, one can define a \emph{4-dimensional effective Planck constant} as:
\beq \label{Planck_constant_1}
\hbar_{eff}=\hbar\bigg(1+{Tr[\theta\eta]\over4\hbar^2}\bigg)~.
\eeq 
Furthermore, it is clear that the commutation relation between coordinates and momenta is not diagonal, 
the off-diagonal elements are proportional to products of $\theta^{\mu\nu}$ and $\eta^{\mu\nu}$. 

We have considered in Ref. \cite{Bertolami_1} the linear transformations Eqs. (\ref{linear_1}) 
to determine the noncommutative Hamiltonian for a particle moving in the $xy$ plane subject to a 
uniform gravitational field $\mathbf{g}=-g\mathbf{e_x}$. As the effects of  phase space noncommutativity are 
presumably small,
we have compared the leading order corrections to the commutative Hamiltonian with the experimental results obtained by 
Nesvizhevsky \emph{et al.}\cite{Nesvizhevsky} for neutrons in the GQW 
formed by a horizontal mirror and the Earth's gravitational field. This has allowed us to set an upper bound of about meV/c 
for the fundamental momentum scale introduced by noncommutativity, $\sqrt{\eta}$. Supposing 
that the fundamental noncommutative length scale, $\sqrt{\theta}$, is smaller than the neutron size, that is, 
of order $1~fm$, we have found that $(\hbar_{eff}-\hbar)/\hbar\lsim O(10^{-24})$. Therefore, the difference between 
the effective Planck's constant and the usual Planck constant has no practical effects.

On the other hand, it has been recently claimed in Ref. \cite{Zhang_3} that for a particular choice of the 
value of the scaling factor $\zeta$ the commutation relation between coordinates and momenta is exactly 
defined by the usual Planck constant. However, it can be shown that 
NCQM models with or without corrections to the value of Planck's constant are, in fact, physically equivalent, 
differing only in the way one defines the noncommutative parameters  \cite{Bertolami_2}. 
In what follows we shall present our analysis of the GQW in the context of the NCQM.

\section{The Gravitational Quantum Well}

We consider now a particle of mass M moving on the $xy$ plane in a uniform gravitational field $\mathbf{g}=-g\mathbf{e_x}$. 
When a horizontal mirror is placed at $x=0$, a GQW is set up. In the direction 
transverse to the gravitational field, $y$, the particle is free, exhibiting a continuous energy spectrum. 
In the direction of the gravitational field, $x$, the particle exhibits a discrete energy spectrum. 
Its wave function in the $n$-th quantum level is given by the Airy function $\phi(\xi_n)$, 
where $\xi_n=(x-x_n)/x_0$ and $x_0\equiv(\hbar^2/2M^2g)^{1/3}$ \cite{Landau}, with energy eigenvalues given by:
\beq \label{GQW_2}
E_n=-\bigg({Mg^2\hbar^2\over2}\bigg)^{1/3}\alpha_n~,
\eeq
where $\alpha_n$ corresponds to the $n$-th zero of the Airy function and $x_n=E_n/Mg=-x_0\alpha_n$ 
to the maximum height that is classically compatible with a particle with energy $E_n$.

The probability of finding the particle is non-vanishing for all values of $x>0$. 
However, when $x$ exceeds the value of $x_n$ for each quantum state $n$, this probability decays 
exponentially, but the particle has a finite probability of penetrating 
a classically forbidden region through quantum tunneling.

These ideas have been used to study the spectrum of neutrons in the quantum well of the Earth's gravitational field 
and a horizontal mirror \cite{Nesvizhevsky_1, Nesvizhevsky_2}. An ultra cold neutron beam was employed, 
with a mean velocity $v=6.5 \ \mathrm{ms^{-1}}$, traveling through a narrow slit formed by the mirror and a 
scatterer/absorber placed above it. Therefore, the neutron flux through the apparatus is measured 
as a function of $x$. For $x>x_n$, neutrons in the $n$-th quantum state 
have a small probability of crossing the gravitational barrier and tunnel into the scatterer/absorber. 
This probability is given by $\exp(-4/3\xi_n^{3/2})$ and vanishes as the slit height increases. 
Hence, for $x>x_n$, neutrons pass through the slit with little loss. 
For $x<x_n$, however, neutrons have a $O(1)$ probability of being absorbed by the scatterer 
and so the slit is not transparent to neutrons. Thus flux of neutrons through the slit is given by:
\beqa \label{GQW_4}
F(x)=F_0\sum_n\beta_n\exp\Bigg({-{L\omega_n\over v}\left\{\begin{array}{ll}
e^{\big(-{4\over3}\xi_n^{3/2}\big)}\ ,\phi_n>0\\
1\qquad\qquad ,\phi_n<0
\end{array}\right\}}\Bigg)~,
\eeqa
where $F_0$ is a normalization factor, depending on the incident flux, $\beta_n$ is the relative population 
of the $n$-th quantum level, $L$ is the length of the slit and $\omega_n\equiv(E_{n+1}-E_n)/\hbar$ \cite{Nesvizhevsky_2}. 
From Eq. (\ref{GQW_4}), one can conclude that, for $x\gg x_n$, the flux of particles in 
the $n$-th state approaches its maximum value $F_0$, while for $x\ll x_n$, this flux tends to vanish. Therefore, 
the classical turning points $x_n$ separate two regions where the neutron flux exhibits distinct behaviour 
for each state $n$. By adjusting the experimentally measured flux to the predicted flux given by Eq. (\ref{GQW_4}), 
the values of the two lowest classical turning points were obtained \cite{Nesvizhevsky_2}, 
and these are in good agreement with the quantum mechanical predictions, namely
$x_1=13.7\ \mathrm{\mu m}$ and $x_2=24.0\ \mathrm{\mu m}$.

We mention that the Equivalence Principle in the context of the GQW has been analyzed in Ref. 
\cite{Bertolami_6}.

\section{Noncommutative Quantum Mechanics}

We consider now the discussed GQW , with Hamiltonian given by:
\beq \label{Hamiltonian_1}
H'={p_x'^2\over2m}+{p_y'^2\over2m}+mgx'~.
\eeq
The corresponding noncommutative Hamiltonian can be straightforwardly obtained using the inverse transformations 
of Eq. (\ref{linear_1}), for $\zeta=1$, 
\beqa \label{linear_inverse}
x'=C\bigg(x+{\theta\over2\hbar}p_y\bigg)~,\ \ 
y'=C\bigg(y-{\theta\over2\hbar}p_x\bigg)~,\ \ 
p_x'=C\bigg(p_x-{\eta\over2\hbar}y\bigg)~,\ \ 
p_y'=C\bigg(p_y+{\eta\over2\hbar}x\bigg)~,
\eeqa
to replace the commutative variables by the noncommutative ones:
\beqa \label{Hamiltonian_2}
H&=&{C^2\over2m}\big(p_x-{\eta\over2\hbar}y\big)^2+{C^2\over2m}\big(p_y+{\eta\over2\hbar}x\big)^2+
+mgC\big(x+{\theta\over2\hbar}p_y\big)=\nonumber\\
&=&{C^2\over2m}p_x^2+{C^2\over2m}p_y^2+mgC{\theta\over2\hbar}p_y+{C^2\over2m}{\eta\over\hbar}(xp_y-yp_x)+{C^2\over8m\hbar^2}\eta^2(x^2+y^2)+mgCx~,
\eeqa
where $C\equiv(1-\xi)^{-1}$ and $\xi\equiv \theta\eta/4\hbar^2$.
It should be noticed that
\beq \label{aux_1}
{C^2\over2m}p_y^2+mgC{\theta\over2\hbar}p_y={1\over2m}\bigg(Cp_y+{m^2g\theta\over2\hbar}\bigg)^2-{m^3g^2\theta^2\over8\hbar^2}~,
\eeq
so that the last term is an additive constant that can be subtracted from the Hamiltonian. Through the redefinition
\beq \label{momenta}
\bar{p_x}\equiv Cp_x\qquad,\qquad\bar{p_y}\equiv Cp_y+{m^2g\theta\over2\hbar}~,
\eeq 
the noncommutative Hamiltonian can be written as:
\beq \label{Hamiltonian_3}
H={\bar{p_x}^2\over2m}+{\bar{p_y}^2\over2m}+{C\eta\over2m\hbar}(x\bar{p_y}-y\bar{p_x})
+{C^2\over8m\hbar^2}\eta^2(x^2+y^2)+mgCx-mgC{\theta\eta\over4\hbar^2}x~.
\eeq
Notice that the last two terms correspond to the commutative gravitational potential, $mgx$.
Thus the noncommutative Hamiltonian is given by
\beq \label{Hamiltonian_4}
H={\bar{p_x}^2\over2m}+{\bar{p_y}^2\over2m}+mgx
+{C\eta\over2m\hbar}(x\bar{p_y}-y\bar{p_x})+{C^2\over8m\hbar^2}\eta^2(x^2+y^2)~.
\eeq

The similarity between the first three terms in the commutative and noncommutative 
Hamiltonians is clear, the only difference lying in the redefined momenta 
$\bar{p_x}$ and $\bar{p_y}$. The constant term in the definition of $\bar{p_y}$, Eq. (\ref{momenta}), 
has, however, no physical meaning as it corresponds to a translation of all the 
eigenvalues of $Cp_y$ by the same amount. It does not imply any change in the commutation relations of this operator either. 
Thus, the only physical difference between $p_y$ and $\bar{p_y}$ concerns the factor $C$ and likewise 
for $p_x$ and $\bar{p_x}$, meaning that the kinetic terms in the noncommutative Hamiltonian differ from the commutative ones only by 
a factor $C^2$.


\section{Bounds on the NCQM parameters}

We study now the physics of the Hamiltonian Eq. (\ref{Hamiltonian_4}). 
To first order in the noncommutative parameters $\theta$ and $\eta$, it is given by:
\beq \label{Hamiltonian_5}
H={p_x^2\over2m}+{p_y^2\over2m}+mgx+{\eta\over2m\hbar}(xp_y-yp_x)=H'+{\eta\over2m\hbar}(xp_y-yp_x)~.
\eeq
Notice that as $C=1+\xi+O((\theta\eta)^2)$, to first order in the noncommutative parameters, 
$\bar{p_x}$ and $\bar{p_y}$ are equal to $p_x$ and $p_y$, respectively. We can then conclude that at this order of approximation the
noncommutative Hamiltonian differs from the commutative one by a term proportional to $\eta$. Hence, the 
configuration space noncommutativity does not influence the GQW energy spectrum to leading order.

It is expected that $\eta$ is a small correction at the quantum mechanical level, and so one can treat 
the new term as a perturbation to the commutative Hamiltonian. The shift caused by this term on 
the energy levels of the system is given by the expectation value of the perturbed wave function of the system. 
We note that as the Airy function is real, $\psi_n(x)=\psi_n^{*}(x)$, and hence
\beqa \label{aux_3}
\langle p_x \rangle_n&=&\int_0^{+\infty}dx\:\psi_n^{*}\bigg(-i\hbar{\prt\over\prt x}\psi_n\bigg)
=-i\hbar\Big([\psi_n^{*}\psi_n]_{0}^{+\infty}-\int_0^{+\infty}dx\:{\prt\psi_n^{*}\over\prt x}\psi_n\Big)=\nonumber\\
&=&i\hbar\int_0^{+\infty}dx\:\psi_n{\prt\psi_n\over\prt x}=
-\int_0^{+\infty}dx\:\psi_n^{*}\bigg(-i\hbar{\prt\over\prt x}\psi_n\bigg)=-\langle p_x\rangle_n=0~,
\eeqa
where, due do the presence of the horizontal mirror, we have used that $\psi_n(x=0)=0$ and the normalization of 
the wave function. Therefore, the term proportional to $p_x$ in Eq. (\ref{Hamiltonian_5}) does not yield 
any shift on the energy levels, whatever the expectation value of $y$. 
Thus, the perturbed potential due to noncommutativity is at leading order given by
\beq \label{perturbative_1}
V_1={\eta\over2m\hbar}xp_y~.
\eeq

This is clearly analogous to a potential describing the effect of a magnetic field $\mathbf{B}=B\mathbf{e_z}$, 
where $z$ is the direction perpendicular to the plane, on a particle of charge $q$, where $qB=\eta/2\hbar$. 
This is of course simply a formal analogy with no physical meaning, as particles in the GQW must be 
neutral, as is the case of the neutrons used in the experiment by Nesvizhevsky and collaborators.

The leading order energy correction to the $n$-th quantum state is 
given by the expectation value of potential Eq. (\ref{perturbative_1})
\beq \label{perturbative_2}
\Delta E_n^{(1)}={\eta k\over2m}\int_0^{+\infty}dx\psi_n^*(x)x\psi_n(x)=
{\eta k\over2m}\Bigg[\bigg({2m^2g\over\hbar^2}\bigg)^{-\frac{2}{3}}A_n^2I_n+{E_n\over mg}\Bigg]~,
\eeq
where the integral $I_n$ is defined as:
\beq \label{aux_4}
I_n\equiv\int_{\alpha_n}^{+\infty}dz\phi(z)z\phi(z)~,
\eeq
and $k=\langle p_y\rangle/\hbar=m\langle v_y\rangle/\hbar=1.03\times10^8~m^{-1}$ for the experiment 
described in Ref. \cite{Nesvizhevsky_2}. The values of the normalization factor 
$A_n$ and of the integral $I_n$ were determined numerically for the first two energy levels:
\beqa \label{numerical_1}
A_1= 588.109\ ,\qquad  A_2= 513.489\ ,\qquad  I_1=-0.383213\ ,~\qquad I_2=-0.878893~.
\eeqa
With these values, the leading order corrections to the energy levels are given by:
\beqa \label{energy_corrections}
\Delta E_1^{(1)}=2.83\times10^{29}\eta~~(J)~\qquad,\qquad
\Delta E_2^{(1)}=4.94\times10^{29}\eta~~(J)~.
\eeqa
Finally, requiring these corrections to be smaller or of the order of the maximum 
absolute energy shifts allowed by the experiment, the following upper bounds for $\eta$ are obtained:
\beqa \label{eta_bounds_1}
|\eta|\lsim& 2.32\times 10^{-61}~kg^2m^2s^{-2}~(n=1)~,\qquad
|\eta|\lsim& 1.76\times 10^{-61}~kg^2m^2s^{-2}~(n=2)~.
\eeqa

These values correspond to the following upper bounds on the fundamental momentum scale:
\beqa \label{eta_bounds_2}
|\sqrt{\eta}|&\lsim& 4.82\times10^{-31}\:\mathrm{kgms^{-1}}\lsim0.90\:\mathrm{meV/c}\qquad~~(n=1)~,\nonumber\\
|\sqrt{\eta}|&\lsim& 4.20\times10^{-31}\:\mathrm{kgms^{-1}}\lsim0.79\:\mathrm{meV/c}\qquad~~(n=2)~.
\eeqa

We compute now the energy correction of second order on the noncommutative parameters. 
The second order perturbed potential is the following:
\beq \label{perturbative_3}
V_2={\theta\eta\over2\hbar^2}{p_x^2\over2m}+{\theta\eta\over2\hbar^2}{p_y^2\over2m}+{\eta^2\over8m\hbar^2}(x^2+y^2)~.
\eeq

The terms proportional to $p_y^2$ and $y^2$ do not affect the particle's energy spectrum in the 
direction of the gravitational field and, hence, do not give rise to any shift on the discrete energy levels. 
Thus, the second order perturbed potential reduces to:
\beq \label{perturbative_4}
V_2={\theta\eta\over2\hbar^2}{p_x^2\over2m}+{\eta^2\over8m\hbar^2}x^2~.
\eeq
There are now two second order terms that can modify the energy spectrum of the particle. 
The first one is proportional to the kinetic energy in the direction of the gravitational field; the second 
one is formally identical to an harmonic oscillator with frequency $\omega=|\eta|/2m\hbar$. 
The energy correction due to the first term on the $n$-th quantum state is given by
\beq \label{perturbative_5}
\Delta E_n^{(2a)}=-{\theta\eta\over4m}\int_0^{+\infty} dx\:  \psi_n^{*}(x){\prt^2\psi_n\over\prt x^2}(x)
=-{\theta\eta\over4m}A_n^2\bigg({2m^2g\over\hbar^2}\bigg)^{1\over3}J_n~,
\eeq
where the integral $J_n$ is defined as
\beq \label{aux_5}
J_n\equiv\int_{\alpha_n}^{+\infty}dz\phi(z){d^2\phi\over dz^2}(z)~.
\eeq

We have computed the value of this integral for the first two quantum states, obtaining:
\beq \label{numerical_2}
J_1=-0.383213\qquad,\qquad J_2=-0.878893~.
\eeq

In order to set an upper bound on the value of this correction, we need not only the upper bounds obtained for 
$\eta$ but also an upper bound for the value of $\theta$. Clearly, the latter cannot be estimated by the gravitational 
quantum well experiment. One can resort to 
the bound on the value of the coordinates commutator, obtained in a different context \cite{Carroll}, 
$\theta\simeq 4\times10^{-40} m^2$ (which correspond to 
$\theta\simeq(10~TeV)^{-2}$ for $\hbar=c=1$) or, consider a more conservative point of view 
and argue that the fundamental length scale introduced by noncommutativity should be at least smaller than 
the minimum scale compatible with the quantum mechanical approach to the GQW problem. 
This scale is given by the average neutron size of about $1~fm$, below which the neutron's internal 
structure becomes relevant. With this hypothesis, one can get an upper bound on 
$\theta$ of $10^{-30}\:\mathrm{m^2}$ and, consequently, the following upper bounds on the 
contribution of Eq. (\ref{perturbative_5}) to the energy correction:
\beqa \label{perturbative_6}
\Delta E_1^{(2a)}\lsim&7.83\times10^{-55}~(J)~\qquad,\qquad\Delta E_2^{(2a)}\lsim&1.04\times10^{-54}~(J)~.
\eeqa
As for the contribution of the second term:
\beqa \label{perturbative_7}
\Delta E_n^{(2b)}&=&{\eta^2\over8m\hbar^2}\int_0^{+\infty}dx\:\psi_n^{*}(x)x^2\psi_n(x)=\nonumber\\
&=&{\eta^2\over8m\hbar^2}\Bigg[\bigg({2m^2g\over\hbar^2}\bigg)^{-1}A_n^2L_n+\bigg({2m^2g\over\hbar^2}\bigg)^{-\frac{2}{3}}{2E_n\over mg}A_n^2I_n+\bigg({E_n\over mg}\bigg)^2\Bigg]~,
\eeqa
where the integral $L_n$ is defined by:
\beq \label{aux_6}
L_n\equiv\int_{\alpha_n}^{+\infty}dz\:\phi(z)z^2\phi(z)~,
\eeq
whose values were numerically determined for the first two energy levels:
\beqa \label{numerical_3}
L_1=0.537596\qquad,\qquad L_2=2.15572~.
\eeqa
Hence, 
\beqa \label{perturbative_8}
\Delta E_1^{(2b)}\lsim3.64\times10^{-38}~(J)~\qquad,\qquad\Delta E_2^{(2b)}&\lsim&6.39\times10^{-38}~(J)~.
\eeqa

We can then see that the contribution of the first set of second order terms is negligible 
in comparison with the contribution of the second term, which is itself $7\ (6)$ orders of magnitude smaller 
than the respective first order correction for $n=1$ ($n=2$). It follows that the perturbation approach is reliable 
using the upper bounds obtained for $\eta$ for both quantum states. Clearly, if had we used the bound on 
$\theta$ derived in Ref. \cite{Carroll}, the energy corrections would have been about ten orders of magnitude 
smaller.

Therefore the Nesvizhevsky \emph{et al.} experiment constrains the fundamental momentum scale to be below the meV/c scale. 
Of course, one could expect the fundamental scale to be smaller than this. An increase in the 
precision of the experiment may lead to more stringent bounds on the value of $\sqrt{\eta}$  if the results 
are still consistent with the theoretical predictions. One should take into account, however, 
that the experimental energy resolution is bounded by the Uncertainty Principle due to 
the finite lifetime of the neutron \cite{Nesvizhevsky_1}. The maximum energy resolution corresponds 
to a minimum energy uncertainty:
\beq \label{uncertainty_2}
\Delta E^{min} \sim {\hbar\over\tau}\simeq1.2\times10^{-37}~J \simeq7.4\times10^{-19}~eV~.
\eeq

If the theoretical predictions are confirmed to this precision by the experiment, 
then one should be able to place the following upper bounds on the value of $\eta$:
\beqa \label{eta_bounds_3}
|\eta|\lsim 5.22\times 10^{-67}~kg^2m^2s^{-2}~(n=1)~,\qquad
|\eta|\lsim 2.40\times 10^{-67}~kg^2m^2s^{-2}~(n=2)~,
\eeqa
which lead to the following upper bounds on the value of the fundamental momentum scale:
\beqa \label{eta_bounds_4}
|\sqrt{\eta}|&\lsim& 7.22\times10^{-34}\:\mathrm{kgms^{-1}}\lsim1.35\:\mathrm{\mu eV/c}\qquad\qquad(n=1)~,\\
|\sqrt{\eta}|&\lsim& 4.90\times10^{-34}\:\mathrm{kgms^{-1}}\lsim0.92\:\mathrm{\mu eV/c}\qquad\qquad(n=2)~.
\eeqa
These are the most stringent bounds that may be obtained within the framework of the GQW  and they imply for the effective 
Planck's constant, the $\xi$ correction: 
\beqa \label{xi_1}
|\xi|\lsim 5.2\times10^{-24}\qquad(n=1)\qquad,\qquad
|\xi|\lsim 4.0\times10^{-24}\qquad(n=2)~.
\eeqa

\section{Quantum-Classical Divide}

We turn now to the discussion of the conditions for the transition between the quantum and the classical
descriptions in the context of the GQW. This is a central and recurrent 
issue in Quantum Mechanics. More recently, this somewhat conceptual and philosophical discussion 
has become a quite concrete experimental problem given the impressive new experiments of interference of 
macromolecules \cite{Arndt, Hornberger, Hackermuller}. In these experiments the conditions 
under which the quantum coherence of complex systems is lost have been carefully examined. It is believed that determining the drawing line 
between quantum and classical behaviour may bring new insights on the nature of macroscopic objects which exhibit 
quantum properties \cite{Leggett}. 
In this work, we study the transition between the quantum and classical regimes
in the GQW. In such a system, particles exhibit a
discrete energy spectrum and present a non-vanishing probability of tunneling
into classically forbidden regions. We analyze how an increase in the particles
mass may destroy these quantum properties, so that the system will, at least
from an experimental point of view, behave as its classical analogue. Hence, we
suggest a generalization of the Nesvizhevsky \emph{et al.} experiment 
\cite{Nesvizhevsky_1,Nesvizhevsky_2} with neutrons to more massive
particles, such as atoms and fullerene-type molecules.

The proposed criteria is based on the assumption that the dependence of separation between quantum 
states is a reliable criteria for establishing the quantum to classical divide, at least within 
the experimental resolution. We shall consider atoms and fullerene-type molecules. 

The GQW is a convenient system for testing if a system exhibits a quantum or a classical behaviour, given that its energy 
spectrum depends on the mass of the particles involved. In what follows we shall analyze the consequences 
of increasing the particle's mass and consider its experimental implications.

We point out that, for all particles, the energy spectrum approaches 
a classical continuous spectrum for high energies. This can be shown explicitly for our 
problem studying the asymptotic form of the zeros of the Airy function \cite{Bertolami_3}.
Indeed, one can show that, both $\Delta E_n\equiv E_{n+1}-E_n$ and $\Delta_n\equiv x_{n+1}-x_n$ tend 
to zero as $n\rightarrow\infty$. Also, $\Delta\alpha_n$ is strictly decreasing with $n$, and hence
the largest separation between consecutive heights $x_n$ occurs for the first two quantum states. 

Let us now consider the effects of increasing the particle's mass. From Eq. (\ref{GQW_2}), one can conclude 
that all energy eigenvalues are proportional to $M^{1/3}$, the same occurring for the separation 
between consecutive energy eigenvalues. This is somewhat unexpected, as one assumes that this separation 
decreases with increasing mass. However, one should note that the experimentally relevant quantities are the heights $x_n$. 
One can conclude that these heights and the separation $\Delta_n$, are proportional to $M^{-2/3}$. Hence, for more massive 
particles, it is more difficult to distinguish consecutive levels. If two of these cannot be separated experimentally, 
the quantum states cannot be distinguished and the spectrum appears classical.

Thus, as the largest separation between values of $x_n$ occurs for the two lowest quantum states, 
if the particles mass is large enough so that the experimental error is larger than $\Delta_1$, there will 
be no means of distinguishing two consecutive classical turning points. In this case, the flux of particles through the 
slit will show a classical behaviour, increasing as $x^{3/2}$, where $x$ is the slit height 
\cite{Nesvizhevsky_2}. The conclusion is that classical 
or quantum behaviour depend on the experimental resolution and on the mass of the particles. This can be quantified 
in the following way: for a minimum uncertainty, $\epsilon$, to measure the slit height, one can distinguish 
at least two consecutive quantum states if the particle's mass does not exceed \cite{Bertolami_3}:
\beq \label{limit_1}
M_{max}=\sqrt{{\hbar^2\over2g}}\bigg({\Delta\alpha_1\over\epsilon}\bigg)^{3/2}~.
\eeq
For instance, with the error of $\epsilon=2.5\ \mathrm{\mu m}$ for $n=1$ \cite{Nesvizhevsky_2}, one could distinguish 
the first two quantum states for particles with a mass $M\lsim 8m_N$, where $m_N=1.67\times10^{-27}\ \mathrm{kg}$ is the mean 
mass of a nucleon. Equivalently, in order to distinguish two quantum states of a particle with mass $M=Am_N$, 
one must require as maximum experimental error:
\beq \label{limit_2}
\epsilon_{max}=\bigg({\hbar^2\over2m_N^2g}\bigg)^{1/3}\Delta\alpha_1A^{-2/3}\simeq 10.3 A^{-2/3}\ \mathrm{\mu m}~.
\eeq
Moreover, the Uncertainty Principle places limits on the experimental resolution 
for particles with a finite lifetime. Measuring $x_n$, an uncertainty $\Delta x_n$ is equivalent 
to an uncertainty $\Delta E_n=Mg\Delta x_n$. Therefore, in order to have sufficient 
spatial resolution to separate the first two quantum states of a particle with mass $M=Am_N$, 
its mean lifetime must be greater than:
\beq \label{limit_3}
\Delta\tau_{min}=\bigg({2\hbar\over m_Ng^2}\bigg)^{1/3}{A^{-1/3}\over\Delta\alpha_1}\simeq 6.3\times10^{-4}A^{-1/3}\ \mathrm{s}~.
\eeq
For the Nesvizhevsky \emph{et al.} set up, the minimum lifetime is $0.63$ ms, which is 
6 orders of magnitude smaller than the neutron's mean lifetime of 885.7 s \cite{PDG}. 
The Uncertainty Principle allows a precision up to $10^{-5}\ \mathrm{\mu m}$, meaning that the experiment can be further improved.

Another implication of the increase on the particle's mass is the decrease in the probability of tunneling 
through the gravitational barrier. As already referred to, in the state $n$ this probability is 
given by $\exp(-4/3\sqrt{2g/\hbar^2}(x-x_n)^{3/2}M)$.  This rapid decrease is expected since in the classical limit particles 
with energy $E_n$ cannot be found at a height greater than $x_n$. Naturally, the frequency $\omega_n$, which can be viewed 
as the frequency of the collisions between the particles and the gravitational barrier \cite{Nesvizhevsky_2}, 
increases as $M^{1/3}$, an effect that is not compensated by the decrease of the tunneling probability. Thus, 
more massive particles exhibit a sharper transition between zero and maximum flux, 
turning the transition to the classical limit $M\rightarrow+\infty$ discontinuous. 
This makes it more difficult to observe distinct quantum states.


\section{Finite size effects: Atoms and Fullerene Molecules}

Till now the focus of our discussion has been on point-like particles. For neutrons, the average radius of about $1 ~fm$ 
is $10^{-10}$ smaller than the value of $x_1$ and so it can be neglected. However, 
as the particle's mass increases, the affect of their size must be taken into account.

Let us then examine the size effects on the GQW energy spectrum. Before that we point out that 
although the GQW problem has only been solved for point-like particles, some of the consequences 
of the particles finite size may still be inferred to
from a qualitative analysis. Indeed, let us assume that the effects of gravity on all particles which make part of 
a massive system are much smaller than the effects of the forces which bind them. Therefore, one can focus on the movement 
of the massive particle's center of mass (CM), which behaves as a point-like particle. 
The most salient difference between these descriptions is that the wave function of a composed particle 
exhibits an additional dispersion around the position of its CM. This dispersion is quantified by the particle's 
mean radius, $R$. Therefore, in the GQW experiment, when the slit height becomes of order $x_n+R$, 
the absorber gets sufficiently close to the wave function of a particle in the $n$-th state and 
the latter can be absorbed at $x_n + R$. Thus the transition between the regions 
of maximum and minimum flux occurs in the neighborhood of $x_n+R$ and not at $x_n$, as for a point-like particle. 
If $R\gsim x_n$, this effect must be taken into account.

Another finite size consequence is that massive particles will tend to leave the 
lowest quantum states less occupied. The horizontal mirror implies the wave function of each of the elementary 
particles that constitute the composed one vanishes at the origin. This implies that the composed particle 
must be above the horizontal mirror. Hence, the particle's CM must be at a height larger than its average radius $R$. 
If $R>x_n$, particles in the $n$-th level are only allowed to be in the classically forbidden region, 
where they have a very low probability of being found. It then follows that 
the great majority of particles will tend to be found in quantum states $n+1$ and higher. The more massive the 
particle the more likely it will lie at higher quantum states. This solves an apparent paradox of our previous 
discussion, namely that the growth in the particle mass implies an increase in the separation between its energy levels. 
However, as its size also increases, the particle is more likely be found in the higher quantum levels, 
where $\Delta E_n\rightarrow0$.

One must also bear in mind that, as the whole particle must lay above the horizontal mirror, 
there is a minimum value of the slit height, 
corresponding to the average diameter of the particle. For $2R$ of order $x_n+R$, one is no longer 
able to observe the transition between zero and maximum flux. This occurs when $R\sim x_n$, so that one cannot measure 
the $n$-th quantum state as the particle cannot be found at this level. Hence, one is unable 
to test experimentally whether particles of radius $R\sim x_n$ are not really in the $n$-th quantum state.

We consider now the properties of the particles that may be used to test the quantum to classical transition in the 
GQW. Clearly, these particles must possess the following features:
(i) They must be electrically neutral so that neither electromagnetic effects overlap the gravitational ones
nor that the decoherence of the particle beam is induced. A high level of symmetry in the spatial distribution 
of electrons prevents polarization effects. Having vanishing 
total spin avoids the coupling to external magnetic fields;
(ii) They must have long lifetimes, so to satisfy Eq. (\ref{limit_3}). Thus, radioactive particles are not suitable;
(iii) The horizontal velocity of the beam must be small so to maximize the time particles remain as GQW states.

Given these requirements, atoms are, after neutrons, the natural choice in the mass hierarchy. 
From conditions (i) and (ii), they cannot be in ionized states and must be stable isotopes with valence electrons 
in $s$-type orbitals. Considering that the main contribution to an atom's mass is due its nucleons, $M=Am_N$, thus 
from Eqs. (\ref{limit_2}) and (\ref{limit_3}) one computes $\epsilon_{max}$ and $\Delta\tau_{min}$. The former 
has been plotted in Figure 1 for $A<88$ (lanthanides and actinides are excluded). From these results one concludes 
that the maximum allowed error lies in the range $0.1-10\ \mathrm{\mu m}$. 
The minimum lifetime can be shown to lie in the range $0.1-0.63$ ms.
\begin{figure}[htbp]
	\centering
		\includegraphics[scale=0.21]{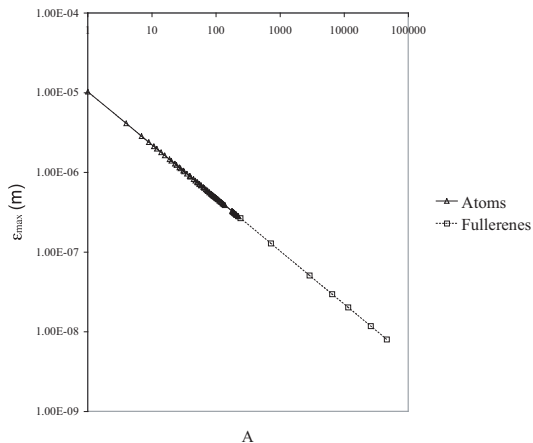}
	\caption{Maximum error for atoms and fullerene molecules.}
	\label{fig:Erro}
\end{figure}

The atomic radius can be estimated using the average value of the radial coordinate for a hydrogen type atom \cite{Gasiorowicz}, 
substituting the atomic number, $Z$, by an effective one, $Z_{eff}$, which accounts for shielding effects 
of inner electrons according to the Slater rules \cite{Slater}:
\beq \label{r_atoms}
R\simeq\vev{r}={a_0\over2Z_{eff}}[3n^2-l(l+1)]~,
\eeq
where $a_0\simeq5.29\times10^{-11}$ m is the Bohr radius and $(n,l)$ denote the atom's valence orbital quantum numbers. 
In Figure 2 the values of the ratio $R/\Delta_1$ for atoms with $A<88$ are shown, and one sees that they lie in the 
interval $10^{-5}-10^{-3}$.
\begin{figure}[htbp]
	\centering
		\includegraphics[scale=0.37]{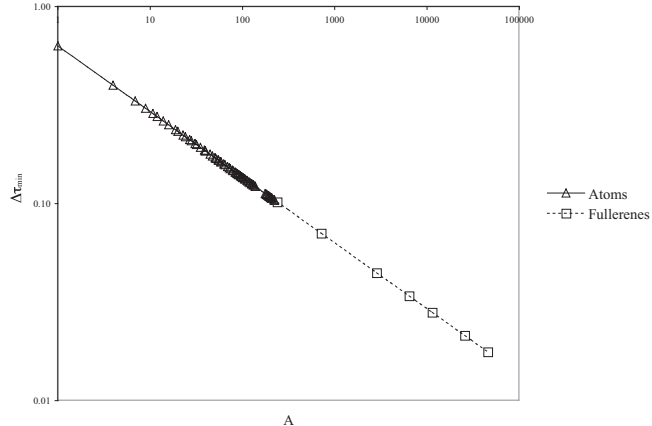}
	\label{fig:Fraccao}
	\caption{Ratio $R/\Delta_1$ for atoms and fullerene molecules.}
\end{figure}

The next scale in the mass hierarchy are molecules. There is a broad variety of molecules, 
but few satisfy our criteria. Among these, fullerenes or \emph{buckyballs}, large groups of $_6^{12}\mathrm{C}$, 
stand out as suitable candidates. These molecules are highly symmetric 
($\mathrm{C}_{60}$ has spherical symmetry) and are globally neutral, although with a slight polarization \cite{Gueorguiev}. 
They are considerably larger than atoms and the higher number of correlated particles enhances the probability 
of decoherence of the beam through interactions with the external environment. Furthermore, they have a large number 
of internal degrees of freedom, rotational and vibrational, and can radiate, yielding in interactions with the 
external environment. Despite these properties, fullerenes are more appropriate than most massive molecules for the GQW study.

For pure $_6^{12}\mathrm{C}$ fullerenes, $A=12N_C$, where $N_C$ is the number of carbon atoms. 
The values of $\epsilon_{max}$ and $\Delta\tau_{min}$ for fullerenes up to 3840 Carbon atoms (after the 
numerical simulations of Ref. \cite{Gueorguiev}) were computed using Eqs. (\ref{limit_2}) and (\ref{limit_3}). 
The values of $\epsilon_{max}$ are exhibited in Figure 1. One obtains that in order 
to distinguish the first two GQW states of these molecules the maximum error must 
lie in the interval $0.01-0.1\ \mathrm{\mu m}$, corresponding to a minimum lifetime of 0.02 to 0.10 ms. 

Supposing a fullerene to be modeled by a sphere of radius $R$, than the area of 
the sphere must be proportional to the number of Carbon atoms \cite{Park}, $R=k\sqrt{N_C}$, 
where the constant $k\simeq4.38\times10^{-11}~m$  is determined using values of Ref. \cite{Gueorguiev}. 
The ratio $R/\Delta_1$ for fullerenes is shown in Figure 2 and is in the range $10^{-3}-10^{-1}$. 
Therefore, one can see that the finite size effects cannot be neglected for the largest fullerenes. 
One can estimate the value of $N_C$ for which $R$ becomes of the order of $x_1$, yielding 
that molecules with more than about $12\:470$ carbon atoms will have a quite small probability 
of being found in the lowest quantum state of the GQW. This means that, the considered fullerenes ($N_C<3840$), 
can be found at the first quantum level.


\section{Conclusions}

In this contribution we have studied the two-dimensional 
GQW. On a first instance we have considered its noncommutative version in phase space 
and used the latest results from the experiment by Nesvizhevsky \emph{et al.} to constrain the 
fundamental momentum scale introduced by noncommutativity to be smaller than $1\ \mathrm{meV/c}$. 
Further improvements in the experimental precision could allow for achieving the minimum upper bounds of about 
$1\ \mathrm{\mu eV/c}$. We conclude that to leading order, noncommutativity in configuration space does not affect 
the energy spectrum of the system. Supposing that the latter introduces a fundamental length scale 
smaller than the neutron's size, we find that the model implies in a modification of the Planck constant which 
is experimentally consistent by many orders of magnitude. 

Furthermore, we have considered the consequences of the increase in the particles mass and size on the energy spectrum of the 
usual GQW. We find that it leads to a decrease in the separation of consecutive classical turning points. 
This implies that the more massive the particles, the greater the precision required to distinguish the lowest 
two quantum states. The precision already achieved allows studying particles up to $A\sim8$. For heavier atoms, 
it is required to increase the precision of position measurements at least by a factor of 10. 
For fullerenes up to $\mathrm{C}_{60}$, this increase in precision may be sufficient, 
but for larger molecules such as $C_{3840}$ an improvement by at least a factor of 1000 is required 

Of course, performing this kind of experiment with sizable particles implies in considerable experimental 
challenges, which involve strong isolation from external agents such as electromagnetic fields, precautions to 
avoid collisions so not to decohere the particle's beam and ultra cold beams of particles to maximize the captured time 
in the GQW. Nevertheless, despite these difficulties, this kind of experiment might be of relevance 
for testing the limits of applicability of the Quantum Mechanics. In the proposed set up, 
the transition to a classical regime is made independently of the phenomenon of decoherence, 
depending only on the mass of the particles in the GQW and on the experimental resolution. 
Observing quantum states of massive particles in a GQW may turn out to be a relevant complement 
to quantum interference experiments with complex molecules. 

\vskip 0.1cm

\centerline{\bf {Acknowledgments}}

\noindent
It is a pleasure the thank our colleagues Cristiane Arag\~ao, Paolo Castorina and 
Dario Zapall\`a for the discussions and for the collaboration which gave rise to part of the work 
reported here.

\end{document}